\documentclass[draft]{agujournal2019}
\usepackage{url} 
\usepackage[inline]{trackchanges} 
\usepackage{soul}
\usepackage{dirtytalk}

\def\apjl{ApJ}                
\def\apjs{ApJS}               
\def\aap{A\&A}                

\draftfalse

\journalname{Earth and Space Science}

\begin{document}

%
%


\title{In-situ Optimized Substrate Witness Plates: Ground Truth for Key Processes on the Moon and Other Planets}

%
%



\authors{Prabal Saxena\affil{1}, Liam S. Morrissey\affil{1,2}, Rosemary M. Killen\affil{1}, Jason L. McLain\affil{1}, Li Hsia Yeo\affil{1}, Natalie M. Curran\affil{1,3}, Nithin S. Abraham\affil{4}, Heather V. Graham\affil{1}, Orenthal J. Tucker\affil{1}, Menelaos Sarantos\affil{1}, Aaron B. Regberg\affil{5}, Diane E. Pugel\affil{4}, Andrew W. Needham\affil{1},  Mark Hasegawa\affil{4} and Alfred J. Wong\affil{4}}

\affiliation{1}{Solar System Exploration Division, NASA Goddard Space Flight Center, , Greenbelt, MD 20711, USA}
 \affiliation{2}{Engineering department at Memorial University of Newfoundland,  St. John's, NL A1B 3X7, Canada}
\affiliation{3}{CREST II/Catholic University of America, Washington, DC 20064, USA}
\affiliation{4}{Contamination and Coatings Engineering Branch, NASA Goddard Space Flight Center, Greenbelt, MD 20711, USA
}
\affiliation{5}{Astromaterials Research and Exploration Science Division,  NASA Johnson Space Center, Houston, TX, 77058, USA}





\correspondingauthor{Prabal Saxena}{prabal.saxena@nasa.gov}




\begin{keypoints}
\item In-situ artificial substrate witness plates we call \say{Biscuits} can monitor processes on surfaces with a low environmental footprint. 
\item We show Biscuits can capture information related to planetary contamination, water transport and material testing/energetic particle flux.
\item The tools are highly versatile and can be a means of assessing key processes by a wide range of groups.
\end{keypoints}

%
%

%
%


\begin{abstract}
Future exploration efforts of the
Moon, Mars and other bodies are poised to focus heavily on persistent and sustainable survey and research efforts, especially given the recent interest in a
long-term sustainable human presence at the Moon.
Key to these efforts is understanding a number of important
processes on the lunar surface for both scientific
and operational purposes. We discuss the potential
value of in-situ artificial substrate witness plates, powerful tools that can supplement familiar remote sensing and sample acquisition techniques and provide a sustainable way of monitoring
processes in key locations on planetary surfaces
while maintaining a low environmental footprint. These tools, which we call Biscuits, can use customized materials as wide ranging as zircon-based spray coatings to metals potentially usable for surface structures, to target specific processes/questions as part of a small, passive witness plate that can be flexibly placed with respect to location and total time duration.    We examine and discuss unique case studies to show how processes such as water presence/transport, presence and contamination of biologically relevant molecules, solar activity related effects, and other processes can be measured using Biscuits. Biscuits can yield key location sensitive, time integrated measurements on these processes to inform scientific understanding of the Moon and enable operational goals in lunar exploration. While we specifically demonstrate this on a simulated traverse and for selected examples, we stress all groups interested in planetary surfaces should consider these adaptable, low footprint and highly informative tools for future exploration.
\end{abstract}

%
%

%


%
%
%
%

\section{Introduction}
The exploration of the Moon and Mars has transitioned to a greater focus on persistent and sustainable survey and research efforts. Multiple countries have now engaged in plans to establish a long term or permanent human presence on the Moon \cite{nasa, esa} and Mars \cite{nasa_2015, jones_2021}. These plans require significant advances in scientific and logistical understanding and as a result, numerous precursor missions have been launched and are planned for the future.  In the case of both bodies, key properties and processes relevant to exploration are known to vary both spatially and over time.  Thus, the longer term goal of establishing a persistent presence on these bodies is dependent on understanding how such processes and properties vary both globally and in the relevant locations for both these worlds. 

Human safety concerns and in-situ resource utilization are two of the most pressing guiding factors in planning for persistent human exploration and habitation of the lunar and martian surface.  A number of studies have stressed the potential hazards that may exist to humans on these surfaces (for a small sample, see \citeA{2016NatSR...629901D, 2012P&SS...74...78R, 2016NatSR...634774P, 2007NW.....94..517H}) and the importance of access to potential in-situ resources \cite{2012P&SS...74...42A, 1559325}. Understanding the present day safety environment and the ability to protect humans on the surface, as well as the ability to access nearby resources, requires localized measurements of a number of key processes and characteristics of the surrounding environment.  Remote sensing and sample analysis techniques are amongst the familiar techniques that can be used to target such measurements, but they are often limited by survivability, logistical, technical and cost constraints. Additionally, they may irreparably modify fragile environments that are both the best targets for a sustainable presence and that hold key scientific information. For example, exhaust heat from exploration activities may modify the extremely cold surface of permanently shadowed regions that are of interest from both a scientific and in-situ resource utilization (ISRU) perspective, and gases given off in exhaust plumes from spacecraft may contaminate otherwise pristine environments. Judicious exploration choices related to choice of landing sites, exploration zones, and how humans live in, travel to and from, and explore these areas on the Moon are likely needed in order to ensure such externalities do not become excessive. Environmentally conscious stewardship of extraterrestrial resources and surfaces that hold promise with respect to scientific content and resource extraction has been recognized \cite{2021SpPol..5701441D, 2014SpPol..30..215K, 2013NewSp...1...60E} as an important imperative in this context and emphasizes the need for low footprint means of exploring key scientific and operational questions.

An additional complication in some of the exploration techniques that may be used for scientific and operational studies is that their measurements may also be relatively ambiguous regarding the differences in details on how key lunar processes are operating at present versus in the the past.  This is due to measurements that obtain cumulative but degenerate information integrated over some fixed time in the past - for example, modification of surface samples by particle flux may be constrained by a number of methods, but those signatures are integrated over time and consequently, chronologically degenerate with regards to total flux over time.  Modeling and additional proxies may be able to intelligently put bounds on the past versus present magnitude of processes, but such an issue also suggests the value of methods that uniquely capture the present day value or magnitude of key processes.  It is with some of these considerations in mind, that we propose the use of in-situ artificial substrate witness plates as a powerful and complementary tool that can be used with more familiar methods to probe key processes in critical areas of exploration for both operational and scientific purposes.

Witness plates are commonly associated with the use of passive sites that are used to record the dynamics and properties of some type of ejecta or local substance of interest, but the term has also been used to broadly refer to the ability of an object to passively record a process \cite{2010IAUS..264..475M, 2012P&SS...74....3C, Saxena_2019, 2019SSRv..215...48M}.  While the use of witness plates to record projectile data has been used in both planetary science \cite{1996LPI....27..475G, https://doi.org/10.1111/j.1945-5100.2008.tb00658.x} and non-planetary science \cite{Singh_1956, doi:10.1177/1464420718759704} studies, the idea of using a fixed passive recorder to observe a process of relevance to planetary and space exploration has been expanded to a number of studies in a variety of other areas. In addition to use in quantifying contamination \cite{Rutherford2018, 2018SSRv..214...19D}, witness plates have also been used to target dynamic processes on the surface of the Moon, including during the Apollo missions. For example, though the Dust Detector Experiment \cite{OBRIEN20111708} on Apollo 11 did use electronic measurements to record the dynamics of the lunar dust environment, those measurements were based upon a passive measurement of the coverage by dust of solar cells placed on the lunar surface. The Apollo Solar Wind Composition Experiment \cite{GeissEtAl}, however, was electronics-free in its measurement and was an entirely passive experiment that was analyzed once it was brought back to Earth. Both of those experiments were able to place novel constraints on key processes that were operating on the lunar surface.  These passive witness plate experiments were finely tuned to measure specific individual processes over a fixed amount of time in manner that supplemented other types of observations. Witness plates have also been used to assess contamination and explore scientific questions related to planetary science beyond the Moon - the Genesis mission \cite{2003SSRv..105..561N} collected samples of the solar wind is a manner similar to the Apollo Solar Wind Composition Experiment and the Perserverance Rover on Mars utilized numerous callibration targets and witness tube assemblies over its duration \cite{2020SSRv..216..141K, 2021SSRv..217....5M, 2022SSRv..218...46F}.

In this paper, we propose a witness plate instrument that builds upon the lessons and utility from these past experiments. We describe how the use of in-situ artificial substrate witness plates (which we call \say{Biscuits} for shorthand) can provide a controlled sample that can be used to constrain the spatially and temporally dependent extant dynamics of a number of key processes relevant to exploration and science on the surface of the Moon.  These Biscuits should be thought of as controlled, calibrated samples that can be used to simultaneously record information regarding a number of different processes over time periods that are critical for exploration and science but that are often logistically difficult to observe.  We use models to demonstrate that, using a preliminary design our team has formulated, these Biscuits can place meaningful bounds on several key processes. Data are extractable through analysis upon return to Earth, through a tailored reactive design of the Biscuit itself, or using future in-situ instrumentation that does not require vast improvement in detection thresholds. Biscuits can serve as key measurement tools that supplement other observational methods, and can do so in a manner that encourages sustainable exploration of the planetary surface in both enabling long term informed exploration and minimizing environmentally damaging externalities of exploration. Since these Biscuits do not necessarily require active electronics, do not require active communication, have low to no thermal footprint, and have a relatively small and adaptable physical dimension and mass, they can preserve a low environmental footprint while providing spatially distributed observations in sensitive areas such as in and around a PSR or around a habitation module for Lunar surface exploration efforts. Since these Biscuits are also highly customizable, substrates can be geared towards operational studies (for example, by exposing different potential building materials) or towards key scientific questions (for example, by using substances that optimally record hydration or micrometeorite flux).  

The remainder of the paper discusses how these Biscuits may be optimized to characterize key processes, how the substrates may be analyzed, and provide case studies regarding potential impact of such tools and issues and considerations that may be of interest.  Section 2 outlines just some of the potential ways in which such an instrument could be designed and explores a Biscuit prototype that our team has developed to demonstrate the potential value of such tools given a number of selected processes.  Section 3 examines these processes and carries out new calculations and cases studies that illustrate that types of constraints that can be put on these processes in more concrete examples.  Section 4 explores the different substrates that may be used in such tools and how they may be optimized, with a focus on one particular process as an illustrative in depth example.  Section 5 discusses analysis of the Biscuits after they have recorded information and how current and potential future detection thresholds associated with returned and in-situ analysis may inform design and strategy in their use - this section again uses illustrative examples versus an exhaustive overview.  Section 6 examines how Biscuits enable sustainable exploration of planetary processes both from a logistical and environmental perspective.  Finally, Section 7 discusses the ramifications of the potential use of these tools, contamination concerns, other complicating factors, and what efforts may be suitable to start to incorporate their usage.

\section{A Biscuit Prototype}

\begin{figure}[t!]
    \includegraphics[angle=0, scale=0.37]{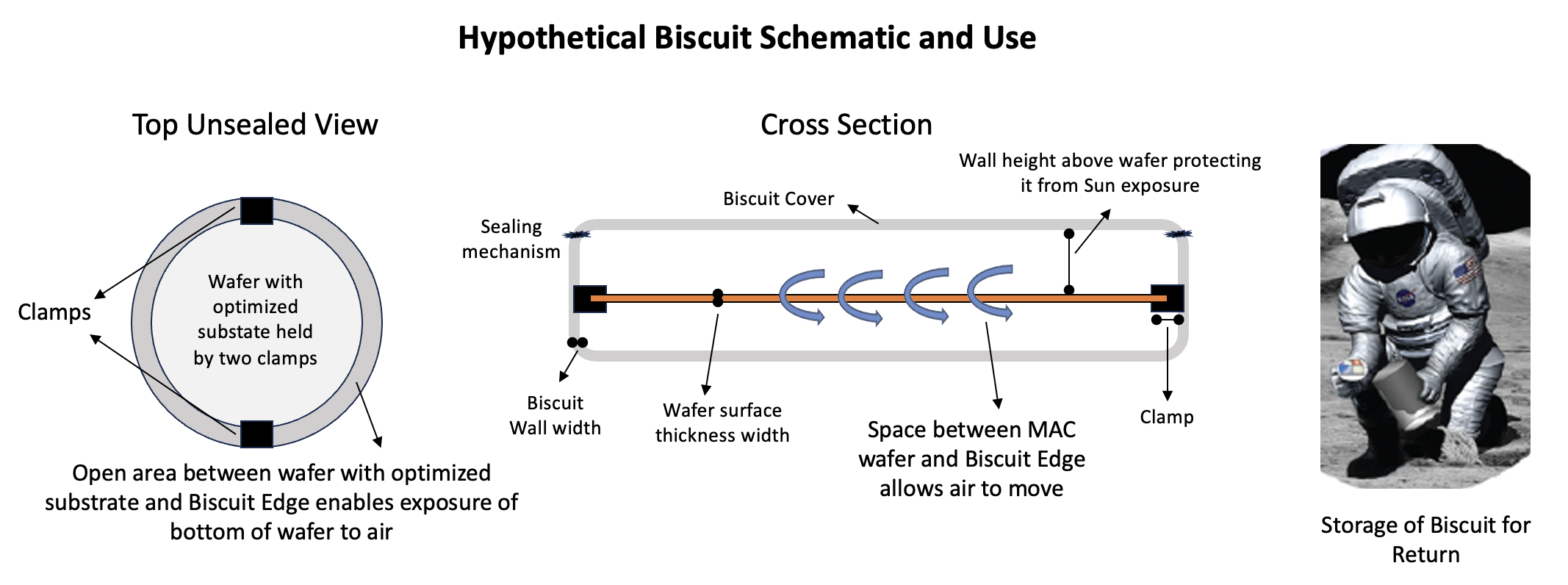}
    \caption{A hypothetical Biscuit schematic, with the housing for the wafer of the optimized substrate(s) and the substrate(s) depicted. The Biscuit would contain a smaller wafer using clamps - the space between the wafer and sides, top and bottom would enable air or volatiles/transient molecular species/ions/exosphere/species of interest to move towards the underside of the wafer. In this hypothetical case, the Biscuit was designed to be sealable using a cover, to have a radius of between 4 to 5 cm, with wafer radius only slightly smaller than that. The roughly 50 to 75 cm$^{2}$ area of the Biscuit would enable easy handling by an astronaut due to its compact size. }
        \label{figure:BiscuitSchematic}
\end{figure}

The Biscuit prototype that guides the case studies and analysis in this study is one that was developed for demonstration purposes in order to explore how such an instrument can put meaningful constraints on multiple key processes simultaneously. At the most basic physical level, the underlying requirements for these Biscuits (including those used for analysis in the case studies) are that they be relatively lightweight and small in dimension so that they can easily be picked up, be passive in not requiring the use of an active power source, and be potentially extensible so that in the future, substrates may be potentially changed.  The motivations behind these characteristics are that these tools must be easily and flexibly placed, be a plausible carry-on for missions that have strict mass requirements, be capable of surviving lunar night in order to track changes over unique and arbitrary time periods, have an extremely low environmental footprint with respect to potentially altering surroundings over time, and be robust enough to capture information in a way that can either be returned to the Earth for analysis or analyzed in-situ.  These characteristics are reflected in a hypothetical Biscuit design schematic, that is depicted in Figure \ref{figure:BiscuitSchematic}.  This schematic shows a hypothetical Biscuit that has a \say{wafer} of optimized substrate(s) material held in place by clamps attached to exterior housing that protects the substrate(s) from the external environment. 
Additional features include a cover that
can be used to close the Biscuit using some type of sealing method, a wall height located some distance above the substrate wafer in order to protect the wafer from Sun exposure related processes, and a gap in between the wafer and housing that can let material access the underside of the wafer. The dimensions for this hypothetical Biscuit are roughly a 4 to 5 cm radius for the Biscuit and a corresponding 50 to 75 cm$^{2}$ area, that both enables a significant enough substrate area for the the case studies discussed later in the paper and that keeps the Biscuit size small enough to be handled easily by an astronaut. Specific processes or experiments that are targeted by a customized design of a Biscuit with a unique substrate may necessitate additional requirements, as is discussed in other portions of this paper, but are supplementary to these core requirements.

Fundamental to the utility of these Biscuits is their use of substrates that are optimized for the collection of materials or phenomena of interest that can be calibrated in a manner consistent with the intended use (for example, tailored selection of substrates based bulk or surface chemical composition for specific binding of desired observable compounds, for specifically chosen maximum concentrations, surface roughness for physical capture of specific size scales, etc.).  The ability to choose a specific amount of a particular substrate in a targeted way and then characterize and calibrate it before deployment essentially allows one to create their own sample for analysis in the future after exposure to the space environment.  These samples are not a replacement for the endogenous samples from a planetary surface that carry invaluable information about the past and present environment they recorded.  However, customizing and characterizing a sample to target it to a particular process over a known time frame enables higher precision measurements that should aid interpretation.  Pre-deployment characterization of the Biscuit through study of the to-be-deployed substrate and/or using an engineering model establishes a control relative to the post-exposure experiment.  Additionally, some Biscuits may also be designed with a built-in control portion.  Depending on the design of such a control, these may also provide either an in-situ or return analysis means of calibrating measurements on the recording portion of a Biscuit. Calibration of the recording part of the Biscuit can also extend to the use of Biscuits themselves as a potential complementary tool to other remote sensing and in situ instruments such a mass spectrometers. While such Biscuits may lack the longer term monitoring and tracking capabilities of endogenous samples that are highly informative about planetary evolution on geological timescales, the ability to tailor them to particular locations, processes and times makes them flexible and highly useful for both scientific and operational purposes.

\section{Capturing Key Processes on the Moon}

\begin{figure}[t!]
    \centering
    \includegraphics[angle=0, scale=0.56]{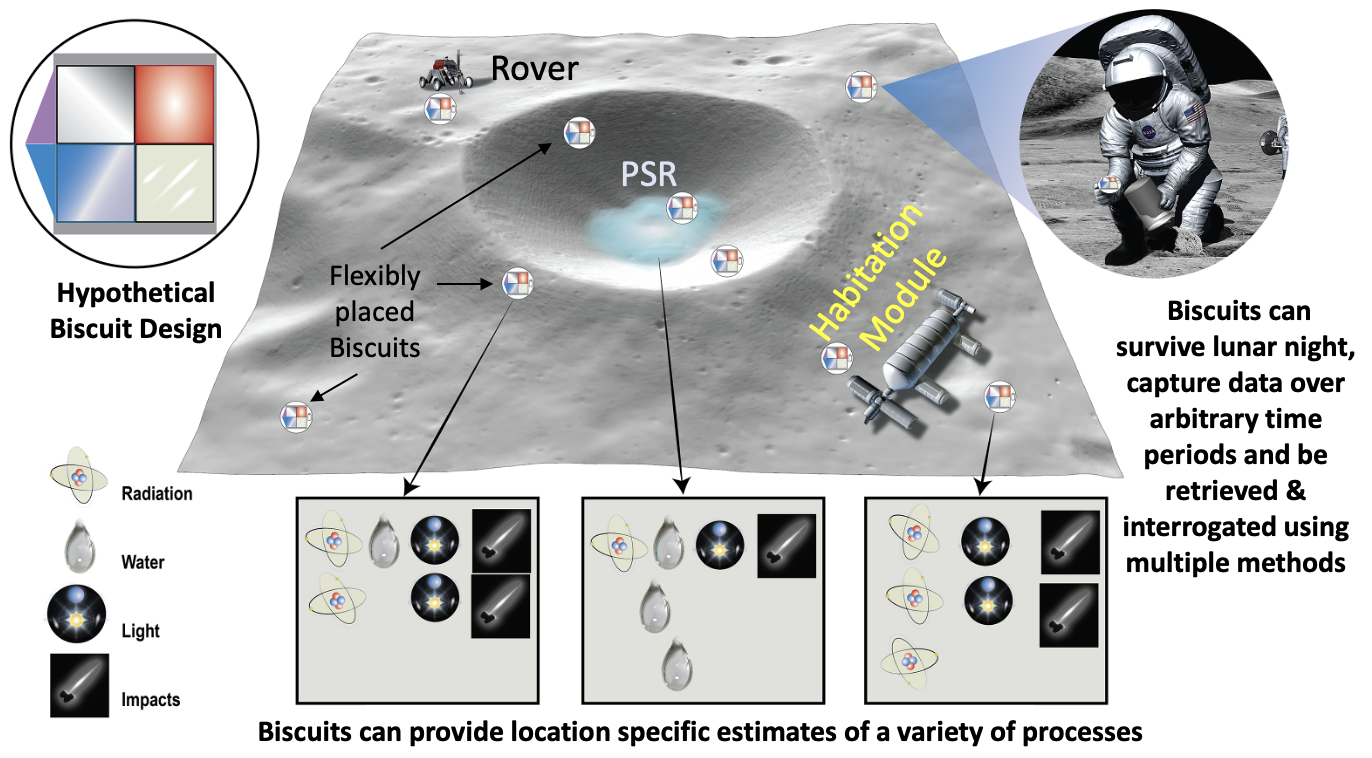}
    \caption{A depiction of the potential usage and some target processes relevant to a Biscuit (energetic particle flux, water transport, illumination and micrometeorite flux are shown here as examples).  The Biscuit cartoon design shown here is meant to illustrate that these tools can be used to track different processes at a number of locations.  The retrieval of a Biscuit by an astronaut and the ability to place them near both a rover or habitation module illustrates that biscuits can be placed and retrieved by multiple methods and can track both scientifically and operationally relevant processes. }
        \label{figure:BiscuitConcept}
\end{figure}

We examined several case studies of how such tools may be able to provide valuable information on processes relevant to science and exploration on the lunar surface.  There are a significant number of areas that could be probed by Biscuits, and as a result these case studies are illustrative, but not exhaustive windows into how they can serve as a key means of exploring critical processes.  These case studies are new and unique examinations of a number of key processes that operate on the lunar surface (including their potential spatial variability) - they use assumptions based upon upon recent literature and current understanding of the underlying processes of interest and may change as that understanding shifts over time.  Note that in each case, the Biscuits used in a particular study were geared towards each substrate.  This demonstrates the customizable nature of these tools, but it may also be possible to capture multiple processes using one Biscuit or to have multiple substrates aimed at several different processes/experiments on a Biscuit.  Examples of additional areas that Biscuits may be useful are discussed at the end of this section.

\subsection{Water Transport and Hydroxylation}

Significant discoveries have been made in lunar science during the last couple decades, including confirmation of the existence of volatiles such as H$_{2}$S, NH$_{3}$, SO$_{2}$, CO$_{2}$ and H$_{2}$O \cite{2010Sci...330..463C} - the existence, abundance, distribution and fractionation of these volatiles can constrain their origin, rates of gain/loss, and processes of transport.  As the most abundant volatile, and most important from a scientific and operational perspective, water has speccifically been of high interest.  Recent discoveries related to water have included water-ice within permanently shadowed polar craters \cite{2010Sci...330..463C, doi:10.1126/science.1187454}, and the presence of mid-latitude hydroxyl signatures in infrared spectra \cite{doi:10.1126/science.1178658, doi:10.1126/science.1179788, doi:10.1126/science.1178105}. On the latter topic, the few percent depth of the mid-latitude hydroxyl 2.8 micron absorption band is consistent with concentrations of hydroxyl at 10-1000 ppm \cite{2017SciA....3E1471L}. It has been suggested that solar wind proton implantation was a possible source for the hydroxyl absorption feature \cite{doi:10.1126/science.1178658, https://doi.org/10.1029/2010JE003711, Tucker2021}. In this case, solar wind protons at 1 keV would implant into the oxygen-bearing regolith, and some fraction of the implanted hydrogen would form hydroxyls. \citeA{Jones2018} has suggested that a fraction of these hydroxyl molecules form surface-escaping water molecules via recombinative desorption of geminal hydroxyls. Recent observations of a 6.1 micron water emission band from the SOFIA observatory at high lunar latitudes \cite{Honniball2020}, suggest some form of trapped water exists near the lunar surface.

The confirmation of the existence of water in permanently shadowed regions on the Moon \cite{2015Icar..255...58H, 2017Icar..292...74F} has been a significant driver of upcoming planned exploration of the Moon and has underpinned plans for a sustained human presence on the lunar south pole \cite{weber2021artemis}.  While the extent and evolution of polar water ice is complicated \cite{LUCEY2022125858, SIEGLER201578}, the existing evidence of water \cite{2018PNAS..115.8907L} in the lunar poles has prompted suggestions that such water may be a valuable in-situ resource that can enable the planned exploration.  A number of studies have examined the extent of potential in-situ resource utilization (ISRU) \cite{2012P&SS...74...42A}, which has included potential resource sites \cite{2020Icar..34713778C, 2022Icar..37714874B} and frameworks \cite{2021PSJ.....2..103L} by which to extract water for ISRU purposes.  However, these polar regions are incredibly sensitive environments and such resource extraction and even just initial reconnaissance ahead of extraction operations may have externalities which damage local environments from both a scientific and future resource extraction perspective. Government \cite{nasa2011nasa} and other stakeholders \cite{MATTHEWS201855, KRAMER2020101385} have recognized this threat and have offered a range of recommendations of how to mitigate such risks \cite{CrawfordManageLunaNP}.  The ability to explore and ascertain ISRU potential of lunar PSRs with a reduced environmental footprint is thus critical for sustainable future exploration of the Moon. Further, it is important to determine if the rate at which water is replenished in such sites is on a time scale relevant to human habitation.

The ability to potentially constrain surface water abundances within a PSR from measurements taken outside of the region would offer an important means of achieving such sustainable exploration.  Recent research \cite{2013P&SS...89...15F, 2015GeoRL..42.3160F, 2019GeoRL..46.8680F} has suggested that water from PSRs can be \say{spilled} to adjacent non-PSR regions through interactions of processes such as incident energetic particle flux \cite{2021PSJ.....2..116N} and impacts within the PSR. These processes liberate and transport molecules from the crater floors that often make up PSRs and move them onto topside regions - resulting in areas of \say{spillage} nearby PSRs.  This spillage model predicts that the transport of volatiles to these adjacent regions depends on a number of factors, but is heavily influenced by the extent and concentration of volatiles within the PSR and the distance from those volatiles.  Crucially, this means that the ability to measure the total spillage of a volatile, such as water, in a region adjacent to PSR may help to probe and constrain key values about the volatile in the PSR region without having to directly sample what may be a highly sensitive region.   

\begin{table}[]
\begin{tabular}{|c|c|}
\hline
\textbf{Included Processes and Factors} & \textbf{References and Inputs}                                                                         \\ \hline
Sputtering Infall and Depletion         & \begin{tabular}[c]{@{}c@{}}\cite{Johnson1990}, \\ \cite{2015GeoRL..42.3160F}\end{tabular}              \\ \hline
UV Photon Stimulated Desorption Loss    & \cite{2015Icar..255...44D}                                                                             \\ \hline
Impact Vaporization \&  Ejecta Infall   & \begin{tabular}[c]{@{}c@{}}\cite{2019JGRE..124..752P}, \\ \cite{2019GeoRL..46.8680F}\end{tabular}      \\ \hline
Thermal Desorption                      & \begin{tabular}[c]{@{}c@{}}\cite{2007Icar..186...24A}, \\ \cite{2019GeoRL..46.8680F}\end{tabular}      \\ \hline
Day/Night Variations                    & Currently 14 days for each                                                                             \\ \hline
Source PSR Size                         & \begin{tabular}[c]{@{}c@{}}Roughly equivalent to a 20 km \\ radius circular source region\end{tabular} \\ \hline
Source PSR Water Abundance              & 1/2\% Water by Weight Abundance                                                                        \\ \hline
Traverse Point Distance from PSR Center & \begin{tabular}[c]{@{}c@{}}For both cases, closest point is \\ 12.5 km from PSR center\end{tabular}    \\ \hline
\end{tabular}
\caption{Polar Water Transport Spillage Model Components }
\label{table:PWTSM}
\end{table}

This hypothesis motivated a case study where we examined how much water that has been transferred to a non-PSR region by spillage can be collected and retained using a biscuit for different lengths of time.  An optimized witness plate would be able to capture a time integrated amount of water spilled from adjacent PSR regions that would be dependent on influx from a particular source a certain distance away versus loss from the biscuit due to removal processes. We examined two different traverses for this case study relevant to potential future Artemis exploration and determined how much water spilled from adjacent PSR regions could be captured along those traverses if Biscuits were left along different points in those traverse for a 10 year period.  The following case discusses the assumptions and details of this model and the values the biscuits would measure along the two traverses for two types of biscuits.  

\subsubsection{PSR Traverse} 

\begin{figure}[t!]
    \centering
    \includegraphics[angle=0, scale=0.50]{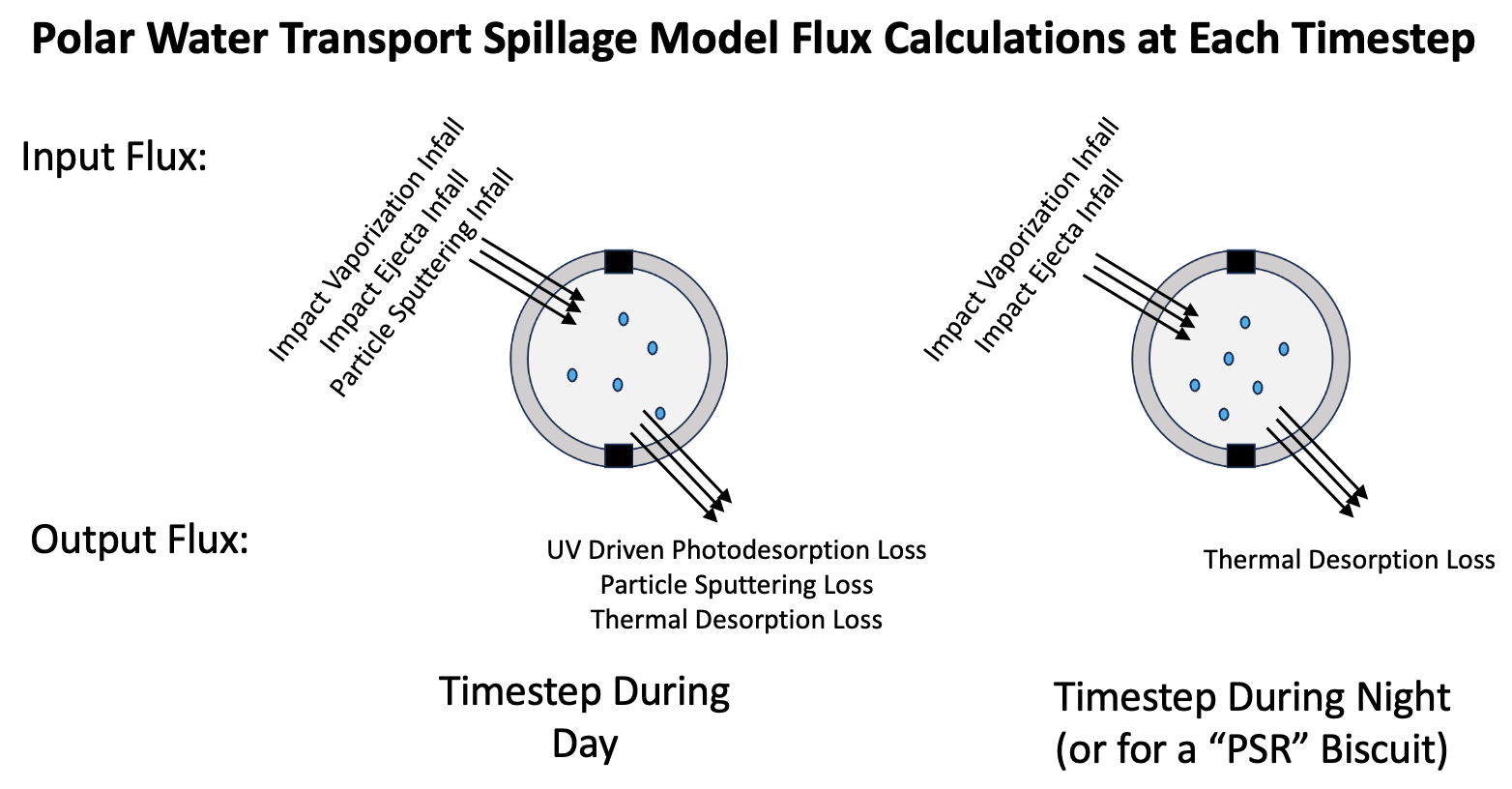}
    \caption{A sketch that shows how the various mechanisms modeled are integrated during a calculation.  During a timestep, input and output fluxes are summed in order to determine the net water transport for a Biscuit, which is then used to calculate a new water abundance for the Biscuit.  These fluxes are dependent upon local properties and vary between day and night.  Night time fluxes are equivalent to fluxes for a \say{PSR} Biscuit, where some of the loss processes due to solar related processes are not included. }
        \label{figure:PSRWTSteps}
\end{figure}

The water spillage model \cite{2013P&SS...89...15F, 2015GeoRL..42.3160F, 2019GeoRL..46.8680F}  that underpins this case study finds that surface processes can continuously release and transport volatiles in PSRs to regions directly adjacent to and above PSRs - with the rate of transport dependent on liberation and transport processes as well as abundance and morphology of water ice in a PSR.  This spillage model considers surface energization processes that act on water in a PSR such as impact vaporization, impact ejection, sputtering and thermal desorption, with additional details given in the references for the model.  Analysis of how much of this water spillage a Biscuit in a region adjacent to a PSR may capture is modeled in this study by a similar assessment of source and sinks for the Biscuit using an analytical PSR Water Transport Model our team has developed. For this PSR Water Transport Model (PSRWT), the net flux of water for a Biscuit can be broken down into a balance between a number of inputs that constitute the total spillage from adjacent PSR regions versus loss processes that remove water from the Biscuit.  The water input processes we consider are taken from the loss processes from the PSR - specifically how much water is delivered by impact vaporization spillage, impact ejection spillage, sputtering spillage and thermal desorption spillage at a particular location from a source PSR.  For simplicity we only consider one source region, whereas a particular location may receive water spilled from multiple PSRs. Loss processes include these same processes that drive spillage from the PSR, but instead are water concentration dependent mechanisms that can remove water from a Biscuit.  These input and depletion mechanisms of water transport for a Biscuit are calculated by incorporating analytical expressions that capture the net flux due to a particular process.  These processes are dependent on location, time and in some cases, concentration specific properties of a Biscuit. The model analyzes the total water content of a Biscuit at a specific location by calculation of fluxes at each timestep while incorporating factors such as day night variation, location dependent temperature and UV/particle flux properties and adjusted total pre-existing abundance of water in the Biscuit.  At the end of a run, the final water concentration of a Biscuit (for example, those given in Figure \ref{figure:WaterTransport}) is an iterative calculation of the changing abundance in the Biscuit as it encounters each successive net flux. Additional detail on these specific processes is given in the the following paragraphs, but a general summary of the processes included during a timestep during both the day and night are given in Figure \ref{figure:PSRWTSteps}, with references and inputs for those processes given in Table \ref{table:PWTSM}.   

The small size of Biscuits are a key to determining the relative magnitudes of the source/sink processes that drive our calculations of the net concentration of water in a Biscuit - in our fiducial case they are assigned a 10 cm x 10 cm (100 cm$^{2}$) size.  This is especially important because in addition to allowing easy physical handling of a Biscuit, this enables an assessment of the magnitude of water loss from a Biscuit due to micrometeorite impact driven processes - one that suggests that impact process driven loss from a Biscuit can be ignored, given its relative contribution versus other processes.  We analyzed this by using estimates of impacts based on \citeA{2009SoPh..256..463M, 2015GeoRL..42.3160F}.  Based upon those two studies, a 100 meter by 100 meter crater floor area on the Moon receives roughly 3.3x10$^{2}$/3.3x10$^{3}$/3.3x10$^{5}$ impactors of mass 10$^{-7}$/10$^{-8}$/10$^{-10}$ kilograms respectively, over a 10 year period.  For a 100 cm$^{2}$ Biscuit, this means that over a 10 year period, the Biscuit would receive roughly 0.3 impacts of micrometeorites with masses of approximately 10$^{-10}$ kilograms, and significantly less impactors of larger mass.  Given that \citeA{ 2015GeoRL..42.3160F} baselined approximately 10 water particles released for each 10$^{-10}$ kilogram incident micrometeorite (assuming a velocity of roughly 10 km/s), we find that based upon this analysis, impact driven processes are an inefficient loss mechanism of water from Biscuits over a 10 year or shorter period and can be ignored in our study.  Due to this finding, we ignore the following impact driven loss processes of water from a Biscuit: impact vaporization, impact ejecta, and impact driven loss from recombinative desorption.  For the last process, since we can choose the substrate used in a Biscuit and given typical temperatures in PSR adjacent regions, a (for example) fused silica substrate Biscuit wound not reach the roughly 600K temperature required for recombinative desorption loss \cite{Jones2018} without impact driven spikes in temperature.

With this simplification, we can start to assemble a balance for the net flux of water at a Biscuit by considering input and loss sources.  Input sources are due to spillage from the adjacent PSR and are driven by impact vaporization, impact ejecta and energetic particle sputtering. Loss processes for water from a Biscuit are thermal desorption, energetic particle sputtering loss and ultraviolet radiation driven photon stimulated desorption.  The balance between these processes at a Biscuit over time is what underpins the PSRWT model.  These input and loss processes are given in the first four rows with their corresponding references and input in Table \ref{table:PWTSM}. The impact driven and energetic particle driven sputtering input processes that add water to a Biscuit are based upon the distance dependent spillage estimates calculated by \citeA{2019GeoRL..46.8680F} and updated using \citeA{2019JGRE..124..752P} for the impact processes and by \citeA{Johnson1990, 2015GeoRL..42.3160F} for sputtering.  Note that in these and other input spillage cases, the total distance dependent transport is based on assumptions regarding the total PSR area and water concentration at the surface - we scale and extrapolate our values based upon the size of the PSRs we select, several choices for the water abundance (1\%/2\% for the two cases in Figure \ref{figure:WaterTransport}), and the total distance from the PSR center. Thermal desorption and UV photon stimulated desorption driven spillage are neglected due to the low temperatures and lack of exposure to the Sun in PSRs.

For water loss from a Biscuit, we use a specific methodology for each of the relevant processes.  In the nominal case of Biscuits that are designed so that the substrate may be exposed to the Sun at times, we include all loss processes. In this case, thermal desorption loss is based upon temperature dependent estimates of the total loss of molecules per m$^{2}$ using \citeA{2007Icar..186...24A, 2019GeoRL..46.8680F} and location dependent temperatures at a particular position using \citeA{WilliamsEtAl2019}. For UV photon stimulated desorption, we determine exposure cross sections based upon Biscuit water abundance using Figure 3 and a photodesorption yield function based on equations 11 and 12 of \citeA{2015Icar..255...44D}.  We attenuate the values based upon the latitude of the location we examine versus the UV photon flux ($>$8eV) of 1x10$^{12}$ per cm$^{2}$ s used for a latitude of 
 45 degrees in \citeA{2015Icar..255...44D}. 
 Calculations of the surface abundance of water in the Biscuit are for this process made relative to the 80 ppm concentration of the sample that \citeA{2015Icar..255...44D} use in their study. Finally, for loss due to sputtering, we use a calculation for sputtering yield based on an assumed mixture of silica and water.  Corresponding binding energies binding energies and proton sputtering yields are taken from \citeA{Johnson1990, 2015GeoRL..42.3160F} (we leave more advanced yield predictions using tools such as SDTrimSP for future work).  The assumed proton flux that drives the sputtering is taken as 2E12 per m$^{2}$s based on \citeA{2019GeoRL..46.8680F} and is not attenuated by latitude for the purposes of this study (so this represents a nominal upper limit on loss rate due to proton flux).

In addition to the baseline Biscuits that can have the substrate exposed to the Sun, we also evaluated \say{PSR Biscuits}, Biscuits that are constructed so that the surrounding casing for the substrate is elevated in order to create an artificial permanently shadowed substrate region. Due to the low maximum elevation angle of the Sun in polar regions, the height of these surrounding regions of a Biscuit does not need to particularly high and is relatively easy to create.  The advantage to this configuration is that the shielding of the substrate region from the Sun effectively eliminates Sun-driven loss processes such as UV driven photodesorption and energetic particle sputtering. The model tracks and flags both the use of a PSR Biscuit and when a Biscuit is in a PSR region in order to avoid confusion between real and modified PSRs. The creation of these artificial PSRs is an analogue to what are likely to be naturally existing smaller scale PSRs created due to surface topography and roughness at many scales in otherwise sun-exposed regions of the south pole - with the key difference being that PSR Biscuits can be placed for specific and cutomized lengths of time and at user defined beginning and end intervals.  Conversely, natural micro-PSRs hold time integrated records of lunar surface processes based on their formation dates and corresponding fluxes.  However, in either case, the use of multiple, nearby micro-PSRs at varying distances from a larger PSR may be able to be used jointly in order to infer larger PSR properties.

\begin{figure}[t!]
    \includegraphics[angle=0, scale=0.44]{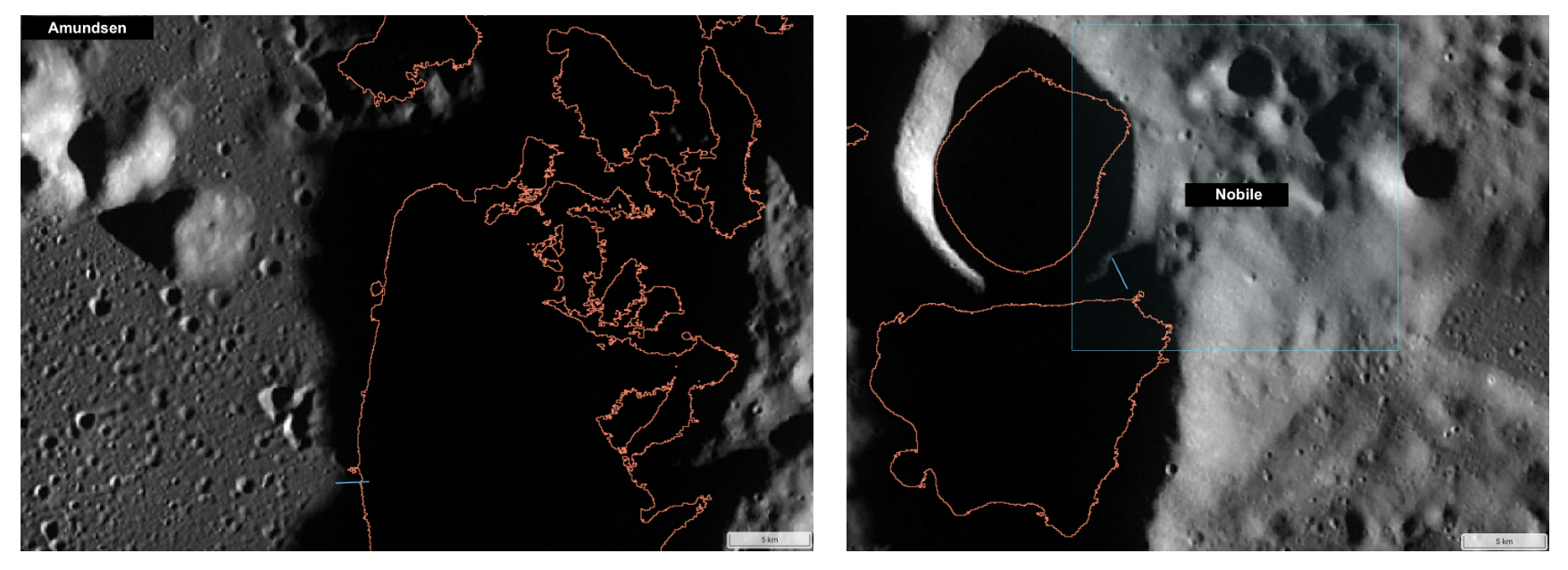}
    \caption{Notional 2 km traverses (blue line segments) within Amundsen crater and the Nobile rim used to model water spillage from PSRs and adsorption onto Biscuits}
        \label{figure:TraverseImage}
\end{figure}

With these input and output processes specified for different target Biscuits, the model carries out a time integration for a point along a selected traverse in a given location in the lunar south pole. For the purpose of this study we chose one day time steps, and used 14 day alternating day and night periods for a full lunar day.  This simplification is incorrect for the lunar south pole as much of the region experiences substantially less periods of time with exposure to the Sun \cite{2011Icar..211.1066M} - this produces underestimates of the total trapped accumulated water concentration in a Biscuit and a more rigorous incorporation of location dependent day/night exposure is left for future work. Nighttime flux for a Biscuit is solely made of incoming water due to impact vaporization and ejecta infall, as all other processes effectively turn off due to no exposure to the Sun and low temperatures. Daytime flux at a Biscuit includes all the processes detailed before. Flux in a PSR region is similar to nighttime flux, as is the use of \say{PSR Biscuits}, with the exception that there, loss due to thermal desorption is included at the location specific temperature.

For this study we selected two 2 km traverses located in the Amundsen crater floor and the Nobile crater Rim (within an Artemis III candidate landing site) that are given in Figure \ref{figure:TraverseImage}. These traverses were chosen due to their proximity to a large PSR region that was roughly 35\% the total area of the fiducial PSR region in \citeA{ 2015GeoRL..42.3160F}, and due to exploration concerns such as an attempt to minimize topographical slope over the traverse, and access to a starting position with favorable illumination conditions. Note that the scaled area used for these traverses is an approximation given the differences in adjacent PSR size and morphology versus the circular PSR examined in the fiducial PSR region.  While the non-circular nature of the PSR near the Amundsen traverse and the presence of multiple nearby PSRs near the Nobile traverse means the influx values are somewhat limited approximations, we believe they still provide a valuable estimate of the magnitude of total water that may be collected by Biscuits, particularly given some of the conservative model choices discussed later.
The closest point in both traverses were approximately 12.5 km from the relevant PSR center.  Calculations of time integrated net water concentration for a hypothetical Biscuit were calculated at 0.2 km intervals along this traverse, with location dependent temperature and potential location within a PSR included.

\begin{figure}[t!]
    \includegraphics[angle=0, scale=0.85]{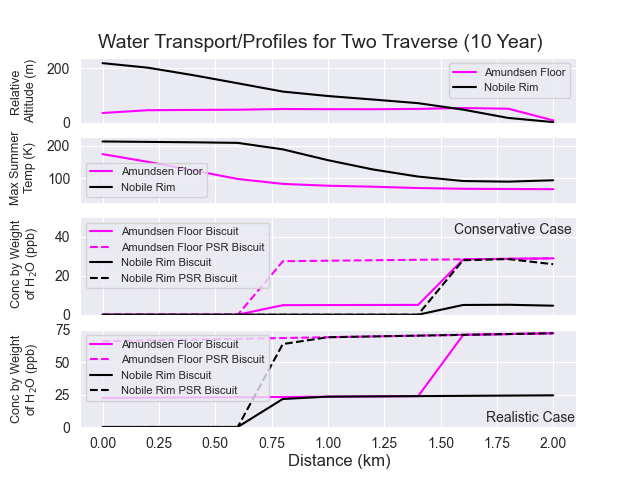}
    \caption{ Results of the PSRWT model in the bottom two panels with contextual information regarding location dependent topography and maximum summer temperature given in the top two panels.  The distinction between the \say{conservative} and \say{realistic} case refers to the treatment of the processes and not necessarily PSR parameters. Differences between the two traverses are heavily driven by local temperature differences.   An additional difference is visible in the Amundsen traverse after $\sim$ 1.5 km, where the traverse enters into a PSR that protects a Biscuit from Sun driven loss processes. The step-like nature of water concentration plotted is due a combination of the granularity of spatial sampling of the Biscuits and the dominance of thermal desorption and sensitivity to key temperatures that are crossed over the traverse. }
        \label{figure:WaterTransport}
\end{figure}

Before discussing the results, it is important to note some of the assumptions we made in this model and the generally conservative effect they had on the estimated integrated water concentration in Biscuits - since this is an initial study to demonstrate the potential of these witness plates to capture information regarding key processes, we left a more detailed incorporation of some factors for future work. A key extrapolation we used was that of the scaled flux using Figure 4 of \citeA{ 2015GeoRL..42.3160F} of impact ejecta and impact vaporization spillage from a source PSR as a function of distances interior to 20 km from the PSR center.  As noted by that study, traverse sites even closer to a source PSR are likely to receive a disproportionate number of lower velocity molecules, and thus a higher flux of molecules than an extrapolation would suggest. Additional choices and assumptions that created lower integrated water concentrations were the previous stated dirunal sun exposure time, the use of the maximum or average summer values for location dependent temperatures and a UV photon flux during the entire day periods.  Similarly, the daytime proton flux used for sputtering loss was alway assumed to be at the zenith angle for simplicity, though the effect on yield is perhaps not as clear in this case.   Using time varying values in combination with additional experimental research work that examines the effect of varying incident angle on both UV flux and energetic particle flux would be room for future improvement. Finally, we did not include any effects from potential local background exospheres.  Estimates based upon the upper limit of 0.62 molecules cm$^{-3}$ for steady-state density in any potential permanent water exosphere \cite{Benna2019} suggest that for a water exosphere to leave any measureable flux onto a Biscuit over the modeled time periods, the exosphere density would have to be at least $\sim$ 4 orders of magnitude greater. Localized water exospheres that exceed this value may be useful to consider in future work.

Results of the PSRWT model are given in the bottom two panels of Figure \ref{figure:WaterTransport}, with contextual information regarding location dependent topography and maximum summer temperature given in the top two panels.  The results in the bottom two panels are for model runs carried out for a 10 year period for both the Amundsen and Nobile traverses and using the two Biscuits types previously discussed.  The distinction between the \say{conservative} and \say{realistic} case refers to the treatment of the processes and not necessarily PSR parameters, though a PSR surface water abundance of 1\% versus 2\% was used in the cases, respectively.  Primary differences between the two cases are that the \say{realistic} case uses an updated value for the impact vaporization and ejecta spillage based on \citeA{Pokorny_2020}, which incorporated the influence of topography and impact of high inclination meteorites and that the \say{realistic} case uses an approximate average summer daytime temperature versus the summer maximum.   

Unsurprisingly, the \say{realistic} case results in larger water concentrations versus the \say{conservative} case. Similarly, the PSR Biscuits, which limit some loss processes, also contain higher water concentration over time.  For these PSR Biscuits, at least some part of the traverse contains regions where a Biscuit would collect a water concentration greater than 20 parts per billion (ppb) over the 10 year period.  This concentration is highly dependent on the distance from the PSR center and the local temperature, which controls thermal desorption (the most effective loss processes for most locations). In locations with temperatures below $\sim$ 100K, water concentration rises substantially for PSR Biscuits - to abundances of approximately 75 ppb. Given the ability to place Biscuits at distances even closer to PSR centers (note that for both cases, the 2 km traverse lengths' closest point to PSR center is $\sim$ 12.5 km, which was chosen for ease of modeling) and the conservative assumptions of the PSRWT model, integrated concentrations of water of several hundred ppb can be plausibly expected in some regions.  The difference between the two traverse locations is almost entirely controlled by local temperature - this is apparent from examining the maximum summer temperature for the two traverses and noting that they roughly correspond to the water concentration points along the traverese.  For example, in Figure \ref{figure:WaterTransport}, the concentration diverges between the two traverses from approximately 0.6 to 1.4 km due to the difference in temperatures (notably, the maximum summer temperatures are on different sides of the key 110K temperature for sublimation of water ice noted in \citeA{2019GeoRL..46.8680F}).  One other difference is the Amundsen traverse after $\sim$ 1.5 km, where the traverse enters into a permanently shadowed region that protects a Biscuit from Sun driven loss processes. A final key note is that the step-like nature of water concentration plotted is due a combination of the granularity of spatial sampling of the Biscuits and the dominance of thermal desorption and sensitivity to key temperatures that are crossed over the traverse. Future work could incorporate more information regarding local PSR properties and greater spatial sampling to help distinguish between the value of competing locations. Notably, the distance dependent profiles also demonstrate that the use of multiple Biscuits at different locations from a PSR and their individual water concentrations may be able to be used jointly to retrieve bounds on PSR properties, such as water abundance. Much of this will be dependent on how such Biscuits could and would be interrogated in order to obtain results, which we discuss in section \ref{sec:sampanalysis}.

The model used for these calculations can be applied to other traverse sites and for a variety of different assumptions.  While it can be used to examine the use of Biscuits or other witness plates, it is notable the the model is agnostic about the nature of the collection site.  Micro-PSRs in otherwise well lit regions adjacent to PSRs should collect water and other materials in similar ways and may hold a treasure trove of information regarding local volatile extent and distribution and past surface processes \cite{Hayne2021}.  The maximum concentration of water that a site can hold is a balance between input and loss fluxes that is dependent on the site size and topography and is an attractive topic for future work. For a water measuring Biscuit, the saturation limit is dependent on a combination of the substrate material and the Biscuit area and design. Multiple nearby small water trapping sites may provide key joint bounds on relatively recent surface processes and volatile delivery as the model predicts that more ancient potential volatile delivery mechanisms \cite{2017E&PSL.474..198S, NEEDHAM2017175} are unlikely to be preserved given flux balances (particularly as impact driven loss becomes more important). Interrogation of multiple micro-PSRs, whether natural or artificial, may be a powerful way to infer lunar south pole volatile content sustainably. Finally, while this model and case study used assumptions and simplifications to demonstrate potential utility of Biscuits, the potential infall of dust and other accumulating material may produce both confounding effects well as scientific opportunity - however mitigation techniques such as placement at varying heights, dust baffles and use of dust mitigating material may be means of targeting specific processes and volatiles.

\subsection{Contamination Concerns and Biologically Interesting Molecules} \label{subsec:Contamination}

Tracking biologically interesting molecules on the lunar surface is an important goal from both a scientific and contamination perspective. The existence, stability and distribution of organic molecules that may carry implications for biology could signal potential habitable environments and/or ingredients that may promote or indicate habitability and are thus of interest to astrobiological studies.  Understanding the presence and extent of such molecules can also be a key means of assessing potential contamination. Both sample handling and prevention of excessive forward contamination benefit from an understanding of the relevant surrounding environment and the molecules they contain that may be relevant to these concerns.

\begin{figure}[b!]
    \includegraphics[angle=0, scale=0.85]{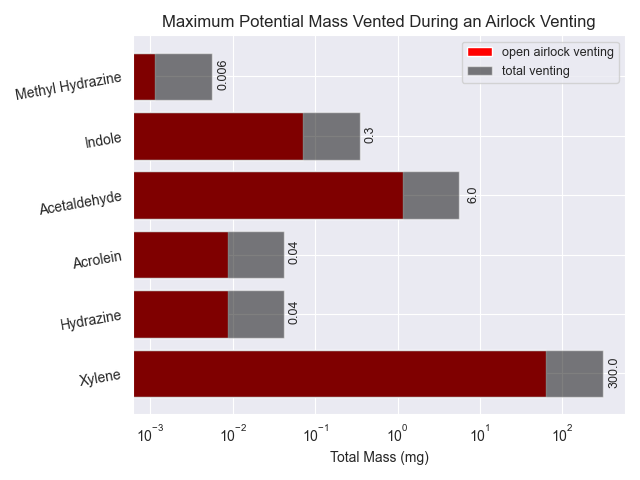}
    \caption{Maximum allowable masses of vented chemicals from an airlock based ISS requirements \cite{wieland1998living}. }
        \label{figure:VentedChemicals}
\end{figure}

Transport of terrestrial microbes and organic material to a planetary body (forward contamination) can both modify an environment that is being explored with exogenous organic signatures and make the assessment of endogenous astrobiological signatures more difficult. Consideration of forward  contamination has been an important concern related to planetary exploration for a variety of planetary targets and while COSPAR planetary protection policies have previously determined that lunar exploration only poses a remote chance that contamination carried by a spacecraft could compromise future investigations,    \cite{doi:10.1073/pnas.061021398, NAP11381, 2020FrASS...7...26N} recent heightened interest in exploration of the lunar south pole has prompted concern regarding the potential of forward contamination modifying ice deposits that may hold key scientific and resource value \cite{2021Natur.589..180W, https://doi.org/10.1029/2020JE006464}. This has prompted a classification of missions to the lunar poles as Category IIb, which requires such missions to record their full organic inventory \cite{fisk2021cospar, CrawfordManageLunaNP}, but does not require those missions to make any effort to minimize or mitigate forward contamination.  Understanding potential vectors for forward contamination, how to monitor and mitigate against contamination from these sources, and how to longitudinally track contamination over time will be key to preventing this unwanted modification.  Artemis represents the first human exploration of a potentially volatile or organic rich planetary environment. It is possible to bake out and otherwise reduce contamination for robotic missions in ways that are not possible with humans in the loop, both due to their metabolic needs and the much larger masses associated with almost every part of the mission architecture. In that context, we examine how Biscuits optimized to detect potential organic contamination sources may be able to be used to assess contamination from a modeled airlock venting scenario during an extravehicular activity, such as those that might be carried out during exploration of the lunar south pole. The upcoming planned Artemis III mission is expected to consist of multiple extravehicular activities over its duration, which also means multiple instances of airlock venting.

The fiducial airlock venting scenario that we model is necessarily based upon past airlock models and biological/contamination limits as it is meant to illustrate the type of contamination that may be of interest to monitor, with a view towards how Biscuits and other tools can be used for that purpose.  The goal of this case study is to determine what plausible masses of air that may be vented during the de-pressurization of an airlock for an extravehicular activity. This broadly can be broken down into two components - determining the total mass of air that is vented from the cabin/space of interest and then applying some type of biological limits on that air in order to use that mass to determine plausible masses of potential contaminants that are vented. 

We use an extravehicular activity on the International Space Station (ISS), where the crewlock section of the airlock is depressurized, as an analogue for potential lunar south pole surface EVAs. While airlocks and suitlocks that may be or have been used in the future \cite{griffin2008lunar, nasa2018office} will have different means of depressurization and protocol prior to an EVA, and consequently will have different vented mass estimates, this scenario gives a real-life analogue we can study that has crewlock dimensions, depressurization details and details regarding contamination limits for air quality requirements all given in citeable literature. Crewlock dimensions, operations, and depressurization details are taken from \citeA{WilliamsISSAirlockVenting} while air quality requirements that we use to estimate maximum potential contaminant masses are taken from Table 32 of \citeA{wieland1998living}. We assume, based upon these references, that the cabin air used in the crewlock during egress will be the sourced from the main cabin air supply, and will thus contain human and mechanical contaminants that are bounded by the prescribed limits.  For the purposes of study, we use 30 day exposure limits, in order to be conservative on maximum limits that would be allowed during the total cabin time for an expected Artemis III mission - where mission duration is expected to be less than 30 days.  Once we calculate total dimensions of the crewlock and use the depressurization schemes to determine the mass vented to space, we can apply the contaminant limits to produce estimates of maximum contaminants vented.

We use an estimate of the total cubic volume for the crew on the ISS as $\sim$8.8 m$^{3}$ based on the Johnson Space Center's ISS airlock chamber\cite{dunbar}. We assume two astronauts in the crewlock for an EVA, each with a cubic volume of $\sim$0.15 m$^{3}$, which yields a total air volume of $\sim$8.5 m$^{3}$. From \citeA{WilliamsISSAirlockVenting}, we take a starting pressure of 14.7 psia, or roughly 1 atm pressure ($\sim$1.29 kg of air per 1.0 m$^{3}$).  Based upon depressurization scenarios, the total pump-down of the crewlock reaches a minimum pressure of 0.5 psia, at which point the extravehicular hatch can be opened and the remaining air in the crewlock is lost to the exterior.  The mass loss from this venting process is $\sim$0.4 kg.  The depressurization from 14.7 psia to 0.5 psia can occur using a variety of different pump and valve combinations for the two steps used during this depressurization.  These different combinations yield different mass loss values to the exterior and require a range of different times for depressurization.  Depressurization from 14.7 to 5 psia can be done in an acceptable amount of time using the depressurization pump while losing no mass to the exterior.  For the reduction from 5 to 0.5 psia, based on the values given in Table 4 of \citeA{WilliamsISSAirlockVenting} and the desirable dwell times described in the reference, we chose the despressurization routine that conformed to timing requirements while losing the least mass possible to space. This was the combination where the depressurizaton pump was used from 5 to 2 psia in combination with the airlock manual pressure equalization valve (which was used during the entire period from 5 to 0.5 psia) and the combination of the Crewlock Manual Pressure Equalization Valve/Vacuum Access Jumper/Airlock Vacuum Access Port (also used from 5 to 0.5 psia).  The total mass loss to the exterior during the depressurization was $\sim$1.4 kg, and the total mass loss to the exterior from start to opening of EVA hatch was \textit{$\sim$1.8} kg.

We were then able to apply the maximum atmospheric contaminant limits from Table 32 of \citeA{wieland1998living} to this mass to produce an estimate of maximum total contaminant mass potentially lost to space in Figure \ref{figure:VentedChemicals}. For the chosen molecules, the total range in maximum vented mass ranged from micrograms to grams.  Our selection of molecules for our estimate in Figure \ref{figure:VentedChemicals} were chosen based upon an assessment of which molecules pose the most risk of confounding the astrobiological or prebiotic chemical record. It did not, however, include molecules which may effect scientific integrity despite being relatively safe with respect to cabin atmosphere contamination, such as water. Indole was chosen because it is a metabolite, while acetaldehyde, acrolein, hydrazine, and methylhydrazine were chosen because they can be associated with possible prebiotic pathways \cite{CleavesII2003, matsumoto1984selective, kolb1994alternative, folsome1981hydrazines}. Xylene was chosen because it is an organic compound that has been found in meteorites and may degrade in the presence of metal oxides \cite{sephton1998delta13c}.  The molecules are a limited demonstration of plausible contaminants of concern that may be potentially tracked by tools such the Biscuits - they are by no means the only molecules that may be of interest and there may be other compounds that are of even higher interest from contamination control, operational and astrobiological perspectives.

The ability to track and measure the possible contaminants as they are potentially left on the lunar surface is a key role that passive tools like Biscuits, as well as other types of instruments, can play. Sampling cabin air that may contain potential contaminants as they are vented out localized areas (such as from external valves during depressurization schemes as depicted in \citeA{WilliamsISSAirlockVenting}) or tracking their dispersal in adjacent areas over time may be a means of constraining true contamination in-situ. Advanced materials that have been used for decontamination of telescope optics and for other purposes, such as the Molecular Adsorber Coating \cite{10.1117/12.964419, abraham2015nasa}, have demonstrated molecule trapping ability at the micrograms per cm$^{2}$ level, including hydrocarbons \cite{abraham2019preliminary}.  This level of trapping using such a substrate could be diagnostic for the plausible contamination levels indicated in the venting scenario, and the incorporation of such substrates into Biscuits such as those described earlier in the paper may be a plausible tool for both the contamination use cases described above.  Both those types of substrates as well as targeted swabbing instruments \cite{2022LPICo2678.2187B} may be effective ways of tracking contamination at multiple stages of exploration and in a variety of scenarios.  

These tracking methods may be especially important given venting and leakage from the types of space suits that may be used on the Moon and other bodies \cite{paul2010requirements, campbell2011advanced, papale2013rapid}, particularly given that humans and associated exploration objects may carry a high microbial burden \cite{10.3389/fmicb.2021.608478}.  Currently, space suits leak continuously during use without any sort of filtration. The nominal leak rate for the US space suits in use on ISS is 100 cm$^{3}$/ min \cite{utc_aerospace_systems_2017}. Many biologically-derived organic compounds could exit the suit alongside vented and leaked gas. Detailed characterization of trace compounds present in the space suit atmosphere could help constrain this potential contamination. Furthermore, portable life support systems are likely to use swing bed technology to scrub CO$_{2}$ from the space suit's atmosphere. Swing beds use reactive amines to adsorb CO$_{2}$ from the air and then release that CO$_{2}$ directly into space \cite{chullen2018swing}. This is a very different system from the Apollo era space suits which used LiOH to scrub CO$_{2}$ from the environment. The contamination potential of these amine based systems is also understudied. 

There are less details regarding venting from future pontential space suits, such as the Axiom space suits baselined for Artemis III and space suits that may be used by other space programs. With the likelihood of higher involvement by commercial vendors and missions by other countries, where NASA may have less control over many aspects related to bio contamination, ascertaining potential contamination using tracking technology will become even more important.  Given the unique sources of forward contamination that may be of concern \cite{10.3389/fmicb.2020.530661}, the apparent present capability that may exist using such tools to track expected levels of contamination described in this case study could be a key means of assessing and mitigating forward contamination on the Moon that can build upon exhaustive efforts that have been used for other targets \cite{2017AsBio..17..363H, 2018SSRv..214...19D}. The lunar research community has the opportunity to learn from studies of exotic environments on Earth, where the interpretation of biological and chemical data would have benefitted from a better understanding of the contamination potential of the research exploration itself. An example might be Antarctica \cite{meyer1962antarctica}, where microbial growth reported in the incredibly challenging conditions of Don Juan Pond \cite{siegel1979life, samarkin2010abiotic} is confounded by windblown contaminants and the pristine record of ice and snowpack is increasingly affected by motorized exploration activities and the presence of crewed field stations \cite{botta2008polycyclic, helmig2020impact}.

\subsection{Solar Activity and Material Testing}

\begin{figure}[t!]
    \includegraphics[angle=0, scale=0.25]{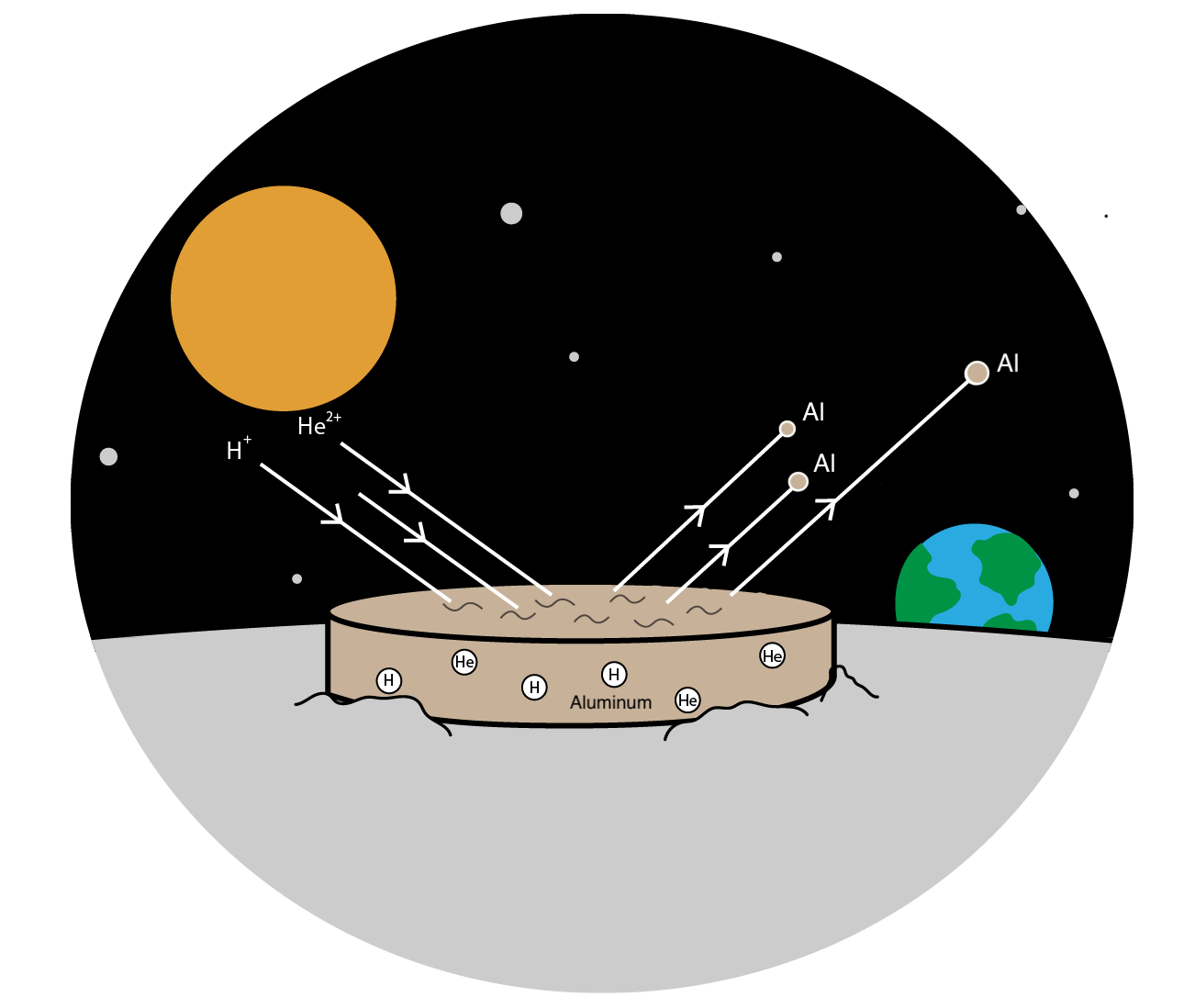}
    \caption{A depiction of a hypothetical material testing/energetic particle flux Aluminum Biscuit.  This image shows how energetic particles from the solar wind can sputter the surface of a Biscuit with an aluminum substrate, embedding those particles, producing damage at different depths, and sputtering off Aluminum particles.  All three of these things can be captured by a Biscuit and can be used to put constraints on time integrated solar wind flux and energy. }
        \label{figure:BiscuitSputtering}
\end{figure}

The interaction of the solar wind (SW) with celestial bodies is a key process to understanding the surface and exospheres of these bodies. The SW is a stream of high-energy charged particles originating from the Sun consisting of electrons, protons, and trace amounts of heavy ions. The term SW encompasses the fast and slow wind along with coronal mass ejections (CME) and solar energetic particles (SEP) events. The energy, flux, and composition of the SW therefore depends on its specific properties \cite{2012JGRE..117.0K02K}. Because airless planetary bodies such as the Moon do not possess an atmosphere or intrinsic magnetic field to shield from the SW; their surfaces are constantly impacted with SW particles. As SW ions impact the surface, they deposit energy, leading to sputtered atoms from the target \cite{2001JGR...10620509K, 2014SSRv..181..121D}. However, there is also a significant spatial variability in these processes with nightside or shadowed regions seeing reduced plasma concentrations of that in the typical slow SW. Furthermore, in magnetic anomaly regions, such as lunar swirls, the incoming ions are deflected, reducing energy and flux by approximately 80\% due to the ambipolar and Hall E field \cite{2015JGRE..120.1893Z, 2016Icar..266..261P, 2015JGRA..120.4719F}.

SW impacts can lead to a number of important surface processes including weathering, damage production in the target, emission of atoms due to collisional sputtering, charging of the surface, and deposition of important volatiles \cite{2015Icar..255..116F, 2019JGRE..124..278T, 2021Icar..35814199J, 2019P&SS..166....9K}. As shown during material tests and simulations for atomic impacts in the low Earth Orbit, sustained high energy atomic impacts can lead to material degradation and damage with limited exposure \cite{2019JSpRo..56.1231M, 2020AdSpR..66.1495M, Banks1989, doi:10.1177/0954008308089705,2022ApJ...925L...6M}. SW impacts therefore represent a significant operational challenge to a sustained anthropogenic presence on the Moon. As such, understanding the role of solar activity on materials exposed on the lunar surface is a critical aspect of designing sustainable human-made lunar structures. Despite the clear importance in understanding these solar processes, further research is needed to quantify their spatial variability and to better explain their effects on various materials.  The Biscuits described in the material testing case study can be designed to simultaneously yield key insight into material efficacy on the lunar surface and energetic particle flux properties - particularly by using multiple, complementarily designed Biscuits.

To demonstrate the unique ability of biscuits to capture valuable solar activity related effects we have conducted a sputtering case study using the binary collision approximation (BCA) simulation tool, SDTrimSP \cite{mutzke2019sdtrimsp}, an extension of TRIM that has been used previously to study SW effects on minerals \cite{2017JGRE..122.1968S, 2018Icar..314...98S, 2020NIMPB.480...10B}[15]–[18]. For this case study we simulate SW impacts onto a pure aluminum target, a commonly used operational material that could be exposed on the Lunar surface for extended durations. Following recommended best practices from \cite{2023PSJ.....4...67M}, we approximate the SW as 96\% 1 keV H+ and 4\% 4 keV He++ impacting an aluminum target with an energy of 1 keV/amu. Using a SW flux of 4x10$^{8}$ cm$^{2}$/s, a total fluence corresponding to 10 Earth years of dynamic SW exposure was simulated. We approximate SW impacts as occurring at incidence angles at 45 degrees from the surface normal. 

\begin{figure}[t!]
    \includegraphics[angle=0, scale=0.65]{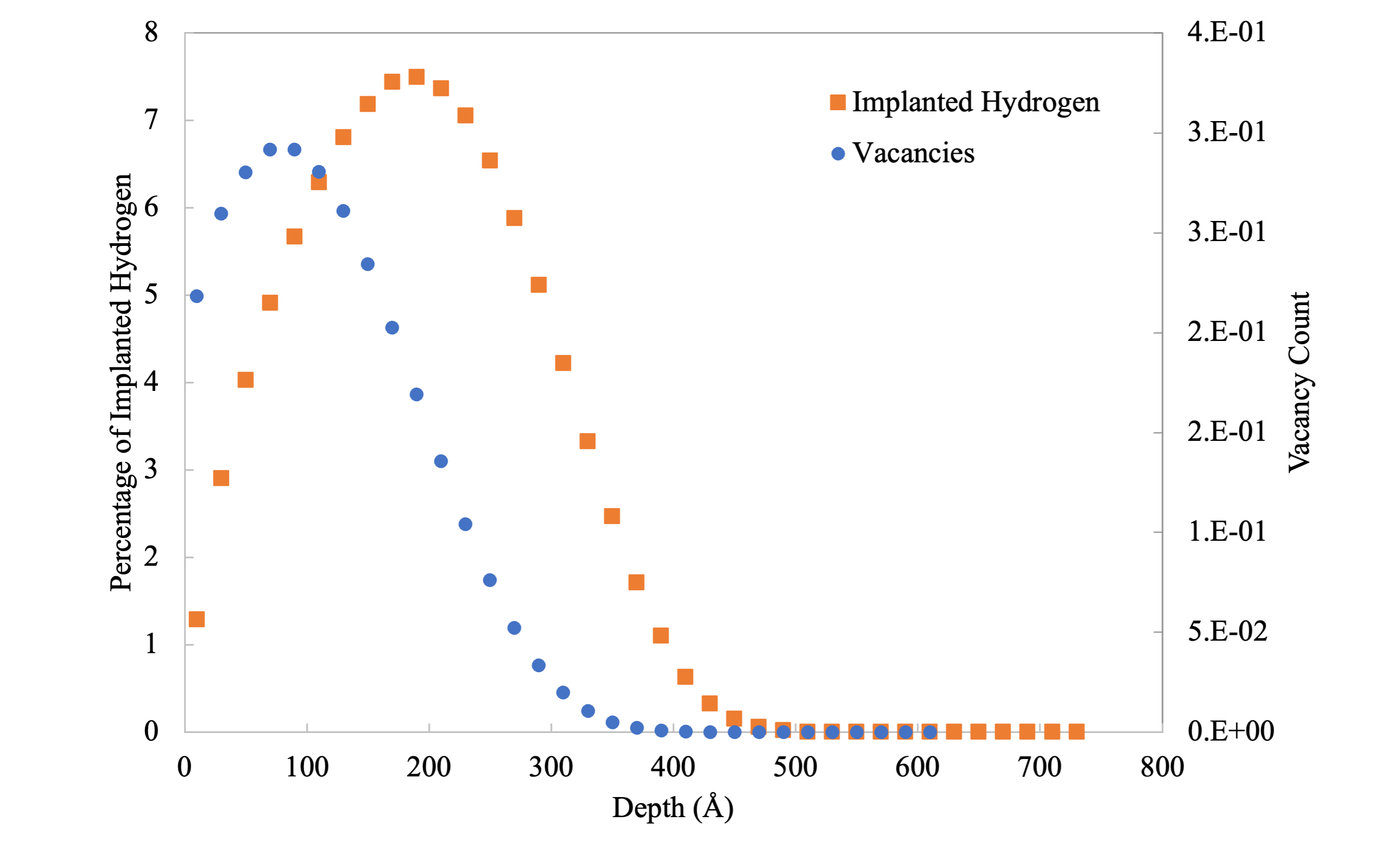}
    \caption{Percentage of hydrogen implanted (orange) (left vertical axis) and number of vacancies per H ion (damage) (blue) (right vertical axis) as a function of the depth for solar wind impacts onto an aluminum biscuit using SDTrimSP.}
        \label{figure:AlumCaseStudy}
\end{figure}

The average simulated aluminum sputtering yield was $\sim$5x10$^{-2}$ Al atoms/ impact. Extending this over 10 Earth years, this corresponds to $\sim$6.3x10$^{15}$ Al atoms/cm$^{2}$. This would correspond to a total mass loss of 0.011 mg and 0.056 mg for 1- and 5-years exposure respectively from a 20 cm$^{2}$ biscuit. This total mass loss from the biscuit is well above detection limits for changes in total mass and demonstrates the applicability for biscuits to be used to quantify SW processes while also assessing the performance of different operational materials. In addition, SDTrimSP can be used to study the depth and damage produced during exposure. As shown in Figure \ref{figure:AlumCaseStudy}, after 10 years of exposure to the SW, significant damage has been created in the sample due to implanted H and He. The depth of this damage, shown as total vacancy count in that depth region, peaks at 150 Angstroms (A) within the sample, shallower than the peak in number of implantations, 50 A. As incoming energetic SW ions make their way through the target they deposit energy along the way, eventually reaching a \say{final} depth. Therefore, the depth of SW induced damage, shown as total vacancy count in that depth region, peaks at 150 A within the sample, shallower than the peak in number of implantations, 50 A. As such, thin, low-mass biscuits can capture the entire deposition profile during exposure.

Overall, these dynamic simulations demonstrate the significant changes that can be induced via SW exposure and highlight the potentially important role these biscuits can play in material assessment. For example, other potential metals, coatings, and oxides could be fabricated into biscuits, exposed for a prescribed duration, and then compared, allowing for a unique in-situ material assessment for future mission design and material selection. In addition, implantation profiles and mass losses from these biscuits can also be used to constrain the integrated flux/spectra at a specific lunar location of interest. Therefore, these biscuits possess both operational and scientific applicability. Like previous SDTrimSP studies, these simulations considered a nominal flux only and did not consider the increased energy that may be experienced due to lower proportion higher mass secondary ions, CMEs, or SEPs \cite{2017JGRE..122.1968S}. As such, these simulations represent a base case of sustained solar wind exposure, where mass loss may be further increased due to other more energetic events.

\subsection{Additional Processes of Interest}

\subsubsection{Dust}

Dust dynamics are an important aspect of the lunar environment. The charging, lofting, and migration of dust grains hold key implications for the evolution of the physical and spectral properties of the Moon and other celestial bodies \cite{WangGRL2016, YEO2021114519}. The deployment of witness plates on the lunar surface can potentially address several open questions on Lunar 
dust dynamics detailed in this section.

The Lunar horizon glow, first observed during the Surveyor 5, 6, and 7 missions in the 1960s is one such phenomenon thought to be closely linked to Lunar dust dynamics \cite{1973ASSL...37..545C, 1974Moon...10..121R, 2007RvGeo..45.2006C}. The presence of a dust cloud above the Lunar horizon raised interesting questions about the possible electrostatic levitation of regolith grains. Part of ongoing debate involves the distribution of dust as a function of height above the Lunar surface. A recent re-analysis of excess brightness measurements from Apollo 15 by \citeA{2011P&SS...59.1695G} using Mie theory revealed steep gradients in the dust scale height and a dust concentration only a fraction of the earlier model \cite{1976LPSC....7.1087M}. It is unclear if the horizon glow observations during the Apollo missions are representative of typical surface conditions as they coincided with major meteoroid activity, which possibly led to increased dust ejections from the surface. It is also possible that Mie theory may overestimate the mass of exospheric dust since realistic lunar grain shapes can be up to several times brighter than their volume-equivalent Mie grains.

Another example of the importance of dust dynamics is the presence of Lunar swirls, regions of high albedo markings that are correlated with localized magnetic fields. They have been proposed to have resulted from dust sorting by near-surface electric fields created by solar wind interactions with Lunar magnetic anomalies. It is possible that the high-albedo markings indicate the presence of finer dust particles, which are relatively brighter \cite{2011Icar..212..480G}. An alternate theory put forth by \citeA{2016Icar..266..261P} posits that strong magnetic fields in the swirl region reflect electrons above the anomaly while allowing protons to enter, albeit at a reduced energy. While their simulations highlight the connection between magnetization profiles and surface weathering, they conclude that in-situ measurements of both magnetic fields and plasma within the magnetic anomaly regions are required to fully understand the nature of the anomalies and the processes involved in space weathering. Biscuits may be a prime candidate to carry out the in-situ dust measurements that could enable greater understanding of these processes and properties.
Finally, characterization of the dust environment would be valuable in determining the utility of the lunar surface for the placement of optical and infra-red telescopes.  Observations from such telescopes on the lunar surface may possess advantages to Earth-based counterparts, but the dust environment is a complicating factor in their efficacy \cite{2023NatAs...7..648S}. 

The development of micro- and even nano-textured materials presents an interesting opportunity to study aggregate dust dynamics passively. Inspired by natural materials such as lotus leaves and rose petals, textures such as micropillars, microdimples, and microchannels have been etched and coated onto surfaces to change their adhesiveness to dust \cite{YILBAS20181, MANDAL2022112311, Patel2019}. Witness plates with varying selectivity towards differing dust compositions or sizes could play an important role in the measurements of dust mobility and migration on the Lunar surface. Simple variations in the positioning, orientation, and height of witness plates would especially shed light on the dynamics and prevalence of dust lofting on the Moon.

\subsubsection{Micrometeorite Mass Flux}

The distribution of impact debris on the Moon is known to be non-uniform, with major sources identified as Helion, Anti-helion, North and South Apex and North and South Toroidal, whose sources are Halley and Oort family comets, Jupiter family comets and main belt asteroids, and dust from Halley type comets, respectively \cite{Janches2021}. Although their radiants are known, their relative intensities are not well quantified due to the fact that the radars used to identify them are not cross-calibrated. Meteoroid flux from asteroidal sources has been widely studied, but recent work \cite{2011ApJ...743...37N, 2011ApJ...743..129N} suggests that meteoroids originating from short period comets dominate the inner solar system mass flux, number flux and total cross section in the micrometer to millimeter range. The velocity distribution, and its effects on the surface are also debated. To obtain the mean asteroidal flux an impact velocity distribution must be assumed. Using the LDEX signal from the LADEE mission, \citeA{2015GeoRL..42.5141S} inferred the relative strength between the Apex, Helion and Anti-Helion sources and found that their relative strengths are variable. This asymmetry has not been seen in Earth-based radars. \citeA{2019JGRE..124..752P} concluded that ~20\% of the asymmetry present in the LDEX measurement is due to unaccounted biases due to the orientation of the LADEE orbit. The study also concluded that the response of meteoroids should be different on the day and night sides of he Moon. The study derived a total flux of MBA meteoroids onto the Moon of 1.4 ton/day, but this is only one of many possible fits to the LDEX data and they have limited latitudinal coverage. The study also found monthly variations of 4 - 5 \% and yearly variations of ~3\%; but with a significant uncertainty of 20\%. Key to tracking both the flux and size/energy distribution of such impacts is that the crater produced by an impact is roughly the same size as the diameter of the incoming meteoroid \cite{2014Icar..238...23S}. Thus, by measuring the crater sizes on Biscuits that are optimized to record micrometeorite impacts, we may be able to derive the size/energy distribution of micrometeoroids, while also adding to statistics on flux.

For example, if the mass flux peaks at about 200 microns (or 10$^{-4}$ g), and 1.4 ton/day is 14.6 g/sec, the mass flux is 3.87x10$^{-17}$ g/cm$^{2}$s or  about 4x10$^{-13}$ particles/cm$^{2}$s. If on the other hand the average particle size is 30 microns, the flux would be about 1.26x10$^{-4}$ particles/cm$^{2}$year. If the biscuit size is 100 cm$^{2}$ then we would expect on average 0.0126 hits/biscuit/year based on MBA meteoroids only. Given that the sources are non-uniform across the Moon, some biscuits would likely have meteoroid hits in one year and others would have one hit in ten or more years. A test of these models could be possible with these types of measurements and a statistically informed concept of operations.

The effect of meteor showers is reported to be large for some showers (e.g. Geminids) but other showers known to have large fluxes have little effect on the exosphere (e.g. Quadrantids) and there are other large temporal increases in the exosphere with no known associated meteor shower (i.e. ~Jan 20, and March 18, 2014 in the LADEE data). In addition to testing the mass and number flux from meteor showers, Biscuits may be able to test latitudinal variations such as those predicted for the various sources.

\subsubsection{Illumination Conditions}

Illumination conditions on the surface of the Moon are central to both exploration efforts and key scientific questions.  Temporal and spatial variations in illumination are significant and often controlling factors in the extent and distribution of water and other key volatiles, the ability to safely enable human exploration of the surface, the ability to sustain and operate instruments and robotic exploration and the ability to remotely observe the surface. Notably, even in regions such as PSRs and other locations with similar low levels of illumination, secondary sources of illumination can have dramatic effects on temperature, radiance and water stability \cite{2011Icar..211.1066M, 2021AcAau.180...25L}.  Coupled with the significant seasonal/temporal and spatial (at multiple scales) variation in all of the sources of illumination \cite{2021AcAau.178..432K, 2021NatCo..12.5607B}, directly and indirectly from the Sun as well as from non-solar sources, this results in a highly variable and relatively limited understanding of true illumination at meter level scales relevant to many exploration and science questions. Tools like Biscuits can be a key means of providing illumination data on spatially and time resolved scales previously not studied.  In addition to windows into a specific time period and place, they can also be used as complementary tools for other, diagnostic exploration of illumination conditions \cite{2018LPICo2087.5028R}. They have the potential to simultaneously assess illumination relevant to exploration and science questions.   One example is characterization of UV radiation, which not only can be responsible for dissociation, ionization and "space weathering" of surfaces, including the production of dislocations in crystalline minerals and other defects, but also can pose a hazard to human health and exploration enabling instrumentation. Given episodic variations in UV radiation and other light due to intrinsic processes and events related to the Sun, such as flares and varying emissions from different regions of the Sun, the use of flexible, long lasting and easily distributed passive tools may be highly advantageous.

\section{Substrate Choice and Strategies} \label{sec:Substrate}

As described earlier, the unique and tailored nature of a chosen substrate is fundamental to the utility of these Biscuits. Not only can choice of material be optimized for the collection of materials or phenomena of interest, but so can calibration procedures, structural and geometric layout, complementary use of multiple substances, and physical properties (for example, size scale, roughness, chemical reactivity).  Flexibility in deploying a particular substrate in a targeted way, with the ability to characterize and calibrate it at different points during a concept of operations, lets a user build their own sample that can minimize uncertainty in the measurement of interest.  Such a range of substrate choices have been used in an analogous way during the Genesis mission \cite{2003SSRv..105..535J}, with substrates ranging from diamond and silicon carbide to multi-layered metal substrates, used to target specific elements. Relevant potential substrates have already been described for some of the more detailed case studies discussed in the previous section.  However, substrate choices are not limited those for the processes of interest - there are a number of substrates that may be useful for capturing or reacting to biologically relevant material and the energetic particle/material testing case study specifically is aimed at testing different potential materials to real environmental conditions.  For other topics such as a micrometeorite flux and illumination state, previously used materials such as aerogel or tune photosensitive substrate may be obvious choices, but are by no means the extent of materials that can be used for those studies.  For the rest of the section we examine potential substrates for the water transport case study as an example of the efficacy of choosing a tailored substrate, and a short discussion of possible alternatives.

Here we describe the scientific potential for a set of silica substrate Biscuits that can provide valuable knowledge of the lunar hydrogen cycle. Returned lunar samples as well as all extraterrestrial samples have inherently been removed from their native environment and transported to Earth. Without great care to isolate these samples from the atmosphere of Earth, which would likely require future technological advances, these samples interact with water and other gases in our atmosphere. A consequence of this interaction is that any analysis must always consider the effects of a potentially contaminated sample, attempt to remove this contamination, and try to ascertain water from hydroxyl in the sample. Interrogating samples along different steps during return encounters the difficulty of spectroscopically distinguishing surface water from surface hydroxyls \cite{2021JGRE..12606845M}, as well as the ability of hydroxyls created from solar wind proton implantation to recombine to form water, and in reverse water dissociating to form hydroxyls. 

However, substrate specificity can be used to untangle and distinguish the processes of lunar exospheric water migration from hydroxyl-forming solar wind proton implantation. One possible example takes advantage of a preheated silica substrate, the process of de(re)hydroxylation of silica as described by \citeA{ZHURAVLEV20001} can be used to distinguish water adsorption from solar wind proton implanted hydroxylation. 
Pretreating the silica substrate for solar wind hydroxylation specificity requires the removal of more than 90\% of the surface hydroxyl concentration. A silica Biscuit pretreated in this way becomes hydrophobic. Migrating water molecules that interact with this surface will not dissociate and form OH. Therefore an increase in the surface OH concentration is solely due to solar wind implanted protons forming OH.  For water migration measurements, a silica substrate that has been heated in vacuum to remove up to 50\% of the surface hydroxyls will dissociatively adsorb water to form surface OH, where the activation barrier for dissociative adsorption is EA = 0. For this silica Biscuit, the cumulative surface hydroxyl content after exposure will be a combination of water dissociated surface OH and solar wind formed OH. These pretreated silica substrates can also be used in tandem to provide a powerful calibration on hydroxyl source.

As described in the case studies, these silica Biscuits can act as both sample collectors, (dust, micro-meteor impacts, volatiles) and are themselves important samples due to surface modification of the substrate via solar wind proton sputtering (oxygen vacancies) and solar wind proton implantation (hydroxylation). It is important to note that this is only one example of a pretreated substrate that can be used to study solar wind hydroxylation and water migration. However, there are numerous other examples that can be employed to measure these processes. For example, an oxygen-free reactive metal surface like sodium, which will react with water exothermally producing NaOH, could be deployed to measure water migration assuming the reactive surface could remain isolated from accidental water exposure.  Other substrates may also be used to assess such processes, and different ones may be used for the same process based upon their efficacy, ease and cost of production, and the methods by which they may be interrogated.

\section{Sample Return and In-situ Analysis} \label{sec:sampanalysis}

Since Biscuit substrates act as both sample collectors, (dust, micro-meteor impacts, volatiles) and are themselves important samples due to surface modification of the substrate, it will be imperative to analyze these samples with various state-of-the-art scientific instrumentation.   Essential to the concept of operations and utility of a particular Biscuit is whether they are required to be interrogated by return to Earth, return to a base of operations or in-situ.  The trade off related to the location of interrogation is one whereby ease and integrity (less likelihood of contamination) of interpretation using an in-situ technique would typically come at the expense of sensitivity or resolution of the measurement due to the expense of carrying more resource intensive, but higher sensitivity equipment to space. For example, in the previous case study related to contamination concerns and biologically interesting materials, the potential use of the MAC substrate may have two different interrogation techniques that reflect this in-situ versus returned dichotomy. While lab analysis  \cite{10.1117/12.964419, abraham2015nasa} of MAC has demonstrated the ability to interrogate its' molecule trapping ability at the micrograms per cm$^{2}$ level, visual identifiable changes in color of the coating due to capture of nonvolatile residue is apparent on the milligrams per cm$^{2}$ level - while both are in the ranges of some of the potential contaminants in Figure \ref{figure:VentedChemicals}, the latter in-situ method is less sensitive.  As a further example of some of the techniques that can be used in analysis of a potential Biscuit, we briefly analyze interrogation methods for space weathering and water/volatile focused Biscuits in the rest of this section.

The geochemical techniques and analysis of space weathering products are continually advancing and improve the limits and range of detection. Solar wind irradiation leads to structural and chemical alterations of the bombarded surface. Distinct differences between fresh and space weathered surfaces can cause optical changes which can be investigated using a combination of microanalytical techniques: transmission electron microscope (TEM), and spectroscopic analysis such as electron energy loss spectroscopy, energy dispersive spectroscopy \cite{1993Sci...261.1305K,1993JGR....9820817P,1993LPI....24...17A, 2007Icar..192..629N}. These techniques have the capability to image the structural and chemical differences withinsubstrates used on Biscuits at the nanometer scale.  For example, microscopy such as transmission electron microscopy TEM or electron microprobes can identify surface defects such as SEP tracks, microcraters, dust accumulation and potentially splotches from deep dielectric discharges \cite{2017M&PS...52..413T, 2018GeCoA.224...64B, 2021M&PS...56.1685K, 2019Icar..319..785J}.  Therefore it will be imperative to analyze these samples with various state-of-the-art scientific instrumentation, and when available, in-situ analysis on the lunar surface. Lunar Missions such as VIPER (Volatiles Investigating Polar Exploration Rover) will be using NIRVSS (Near-Infrared Volatiles Spectrometer system to make measurements of the lunar surface \cite{2015AdSpR..55.2451R}.  Similar IR spectrometers, hand-held X-ray Diffraction, and Laser Induced Breakdown Spectroscopy instruments have all been proposed or flown to the lunar surface could be used to study the evolution of Biscuits over various time frames. 

Elemental and Chemical composition of a Biscuit relevant to water/volatile processes pre and post exposure at the lunar surface can be measured using other microanalytical techniques such as electron microprobe analysis, Inductively Coupled Plasma Mass Spectrometry (ICP-MS) and Laser Induced breakdown spectroscopy (LIBS). The detection limits of these instruments are from below ppb (ICP-MS) to ppm (LIBS).
Infrared (IR) spectroscopy measurements of silica Biscuits relevant to water transport processes can be performed to determine the concentration of hydroxyls at or near the silica surface before and after exposure. Both ATR (Attenuated Total Reflectance) and/or DRIFTS (Diffuse Reflectance Infrared Fourier Transform spectroscopy) techniques can be used to determine the concentration of hydroxyls in the silica substrate as well as other modern FTIR techniques. These techniques can be extremely sensitive to the OH surface concentration depending on the instrument design, where in the mid-IR reports of parts per billion has been reported \cite{doi:10.1366/0003702914337137}.   It has been also described that DRIFTS can be used to identify and determine the amount of low volatility residues (LVRs) on the surface, where in this case we could target forward contamination during the exploration stages \cite{10.1117/12.481658}. This LVRs method detection limit was determined to be 0.01 microgram per cm$^{2}$ for six common organic contaminants.  Key to all these techniques and the overarching context of measuring signatures of interest on Biscuits, is that improvement in technology is continually occurring, in a manner that can be driven by requirements for in-situ or highly sensitive measurements of returned samples.

\section{Sustainable Exploration of Planetary Processes} \label{sec:sustainability}

From a scientific and operational perspective, the case studies we examined suggest that optimized substrate witness plates such as Biscuits can help examine key planetary surface processes in unique ways with respect to location and time.  However, the broader additional advantage they have is the ability to interrogate these processes in a way which may be both less resource intensive with respect to engineering constraints while also mitigating potential negative externalities in the target exploration areas. This is true with respect to exploration areas related to both scientific and human exploration goals.

This low environmental footprint is key for a number of reasons.  Persistent exploration of a region such as the lunar south pole has often included the concept of a base camp of operations.  That location and the nearby areas will naturally be subject to higher levels of potential modification by human exploration.  This in turn will effect the integrity of scientific operations, the potential efficacy of infrastructure development and potentially hazard potential for explorers \cite{CrawfordManageLunaNP}.  Forward contamination and modification of the local area, from both a biological and non-biological perspective, are pathways by which this may occur.  However, changes in the surrounding area that can result in localized niche properties for planetary surface processes are another - for example, payloads in the region may create unique obstacles that produce wakes in energetic particle flow.  Similarly, exploration of the local area may modify surface regolith properties and albedo, which may have significant impact on the dust environment, surface temperature conditions and a range of other properties. Given these potential impacts, minimizing impact on the local exploration area and characterizing the effects of those impacts longitudinally in time is an important standard by which to assure that local properties remain well understood.

The use of Biscuits possesses advantages that can help minimize that impact. The low size footprint and passive nature of Biscuits means they inherently have a low potential modification and waste footprint.  The flexibility of placing and retrieving Biscuits, along with their complementarity to other types of techniques and instruments used to interrogate surface processes, means the marginal impact of and retrieval can be chosen in an optimized way. This optimization is key for two additional reasons.  While operation of Biscuits may be a relatively low impact process, placement and retrieval of Biscuits (if they will not only be analyzed in-situ) may still result in a higher environmental footprint.  At minimum, even if no mitigation steps were taken, Biscuits may still be a more environmentally friendly choice as other instruments or tools would also need to be at least deployed, and would still have operational externalities. However, choosing deployment and retrieval in a more environmentally friendly way (for example by placing or retrieving multiple Biscuits at the same time, carrying out placement/retrieval along with other activities, and/or using robotic or autonomous means of deployment/retrieval) can mitigate the footprint those actions may have and add further advantage to Biscuit use.  Secondly, optimization of placement and retrieval is also critical to mitigating contamination of a Biscuit during those processes.  Again, a well thought out and tailored concept of operations can be implemented to limit potential contamination, as can inherent properties in Biscuit design (such as sealing after exposure period, including remote sealing mechanisms and use of control Biscuits in the intermediate steps).  Supplementary design or implementation techniques can also be highly effective for specific use cases - for example, in a case where a Biscuit is used to capture potential contaminants vented out an airlock, the use of a protective tunnel after placement can serve as a barrier from outside contamination until it is removed.    The relative ease of adaptation, targeting, and manufacturing of Biscuits, along with the potential low cost associated with that, can enable access to a broader range of groups interested in questions related to surface processes.  In addition to groups that may want to assess very specific or spatially/temporally resolved processes or questions for operational and scientific purposes, this low cost can also enable access from groups at different educational levels, including the public.  

For exploration purposes in general, Biscuits may enable distributed analysis of a region that can be incorporated into data infrastructure for astronauts in order to support both real time operations or to feed into forecasting for future operations. Their small size and flexibility means in some cases they can also be tested in analogue environments such as through Earth-based field work, in order to allow for integration and testing prior to use in space.  The use of targeted substrate witness plates such as Biscuits can be a flexible exploration tool in two other significant ways.  They can be used on a variety of different solar system bodies, with an obvious extension of use on the Moon as a proving ground for potential application to use on Mars, where contamination and modification concerns may be even more heightened. Finally, the concept is a relatively simple and universal one - whose use is not confined to any particular space program or geopolitical boundary.

\section{Discussion} \label{sec:Discussion}

The use of targeted substrate witness plates, such as Biscuits, have the potential to make possible a new level of confidence and power in understanding surface processes relevant to exploration and science.  However, there are likely to be numerous issues and challenges that will need to be considered when implementing the use of any specific version of such a witness plate.  In addition to the need to have a well delineated concept of operations for a Biscuit, one such example is the complication of potential accumulation effects - the likelihood that a particular witness plate may receive other material or inputs that are not the specific target of interest. 

For example, in the case of a Biscuit targeted at assessing water transport and spillage, we again use as an example that the Biscuit may also capture dust depending on its placement and design.  Additionally, depending on the design of a Biscuit, there may be other processes that modify the substrate in unwanted ways. A key means of mitigating such effects (and other potential complications) is leveraging the flexibility of Biscuit design and placement. For example, in the case of unwanted dust, the Biscuit could be placed at different heights or for different periods of time to change susceptibility to dust capture.  The substrate or analysis method (including calibration steps given the low expense/mass of Biscuits) could be designed to be more resilient to monitoring the process of interest in spite of dust capture, or design considerations such as the use of a geometry, baffle or anti-dust coating  could be used.  Conversely, the fact that multiple processes may be operating on a witness plate at a particular location may be an advantage, as a multi-section Biscuit with potentially multiple different substrates could be used to simultaneously track those processes.

Pre-characterization of such witness plates would be key to ensuring they are successful.  However, the small size of the Biscuits and their targeted nature means that they may be more easily exposed to multiple testbeds, and that additional calibration versions that can be a means of discerning potential contamination may be left at different stages without the need for a prohibitive amount of resources.  The specificity with which they can be designed and flexibility in placement also means they can be a very low risk, potentially high yield tool that can accompany nearly any type of surface investigation.  The ability to place them on unmanned landers, distribute them from orbiters, have them placed by robotic landers/rovers and to allow placement/analysis by human explorers means they can be an accessible calibration/operational/scientific tool for most types of missions.  While we have several case studies in this paper and have shortly discussed several other potential processes of interest, the range of processes that Biscuits may be able to help probe are enormous: for example, a non-exhaustive list could include processes such as measurement of galactic cosmic ray/solar energetic particle flux, total micrometeorite mass flux, non-water (such as carbon dioxide and methane - key exosphere components) elemental/molecular transport and deposition, local temperature variations, the influence of human induced landing and takeoff activity such as plume effects, ground truth calibration for remote sensing, spatially and time sensitive magnetic field measurements in Lunar swirls and a host of other processes for which they can serve as time capsules. Excitingly, Biscuits may be an optimal way to record very slow/low concentration processes that are impossible to measure during short active mission phases or by any realistic electronic detection system.  Finally, the connection to use on other airless bodies as well as to Mars has already been noted and is fairly straightfoward, however the flexibility and simplicity means that such tools could be modified for use on other targets as well, including Earth. The small, customizable nature of Biscuits means their use cases are only limited by the imagination, while their potential value will be driven by ingenuity.

\section{Open Research}
The traverse and spillage data (including traverse elevation, particle flux and sublimation rate data) used in the PSR Traverse case study was taken from the cited references and is also available (along with the code) at https://github.com/psaxena2/PSRWT (DOI: 10.5281/zenodo.8264883) \cite{psaxena2_2023_8264883}.  Please contact the author for additional information on the code and data.  Code and data are available for use by the public under the GNU General Public License v3.0.  

Additional case studies are based upon the publicly available third party data and code contained within the references that are cited. The PSR source infall and depletion data on which this PSRWT calculations are based is from \citeA{2015GeoRL..42.3160F, Johnson1990, 2019GeoRL..46.8680F, 2019JGRE..124..752P}, UV loss rate data from \citeA{2015Icar..255...44D}, and thermal desorption data from \citeA{2007Icar..186...24A, 2019JGRE..124..752P}. Airlock case study data on depressurization details are taken from \citeA{WilliamsISSAirlockVenting} and maximum potential contaminant masses from \citeA{wieland1998living}.  The SDTrimSP code used for the material testing/particle sputtering case study is available at \citeA{mutzke2019sdtrimsp}.


\acknowledgments
This work was supported by the NASA Solar System Exploration Research Virtual Institute, United States cooperative agreement number 80NSSC20M0021 (SSERVI LEADER: Lunar Environment and Dynamics for Exploration Research). Yeo’s research was supported by an appointment to the NASA Postdoctoral Program at the NASA Goddard Space Flight Center, administered by Oak Ridge Associated Universities under contract with NASA. \\


%
%



\bibliography{agusample}

%
%
%
%
%

\end{document}